\documentclass{aa501}
\usepackage{graphicx}
\def\Msun{M$_\odot$}
\def\Mv{M$_{\rm v}$}
\newcommand{\Teff}{\mbox{T$_{\rm eff}$}}

\def\simgt{\lower.5ex\hbox{$\; \buildrel > \over \sim \;$}}
\def\simlt{\lower.5ex\hbox{$\; \buildrel < \over \sim \;$}}
\def\aj{{AJ}}%
\def\apj{{ApJ}}%
\def\apjl{{ApJ}}%
\def\aap{{A\&A}}%
\def\aaps{{A\&AS}}%
\def\memras{\ref@jnl{MmRAS}}%
\def\mnras{{MNRAS}}%
\def\pasp{{PASP}}%
\def\physrep{{Phys.~Rep.}}%
\begin{document}
   \title{Helium variation due to self--pollution among Globular Cluster
stars:}
   \subtitle{Consequences on the horizontal branch morphology}
   \author{F. D'Antona\inst{1},
	  V. Caloi\inst{2}, J. Montalb\'an\inst{1}, P. Ventura
	  \inst{1}, and R. Gratton \inst{3}%\fnmsep\thanks{Just to
%show the usage of the elements in the author field}
          }

   \offprints{V. Caloi}

   \institute{INAF, Osservatorio Astronomico di Roma\\
	      \email{dantona@mporzio.astro.it}
         \and
	     Istituto di Astrofisica Spaziale e Fisica Cosmica CNR, Roma\\
	     \email{caloi@rm.iasf.cnr.it}
	 \and
	     INAF, Osservatorio Astronomico di Padova\\
	     \email{gratton@mostro.pd.astro.it}
	     %\thanks{The university of heaven temporarily does not
	     %        accept e-mails}
             }

%\date{Received September 15, 1996; accepted March 16, 1997}
\date{}

   \abstract{It is becoming clear that `self--pollution' by the ejecta of
massive asymptotic giant branch stars has an important role in the
early chemical evolution of globular cluster stars, producing CNO
abundance spreads which are observed also at the surface of unevolved
stars. Observing that the ejecta which are CNO processed must also be helium
enriched, we have modelled stellar evolution of globular cluster stars by
taking into account this possible helium enhancement with respect to the
primordial value. We show that the differences between the main evolutionary
phases (main sequence, turn--off and red giants) are small enough that it
would be very difficult to detect them observationally. However, the
difference in the evolving mass may play a role in the morphology of the
horizontal branch, and in particular in the formation of blue tails, in those
globular clusters which show strong CNO abundance variations, such as M13
and NGC 6752.
   }
\titlerunning{Helium variations due to self--pollution and HB morphology}
\authorrunning{D'Antona et al.}

   \maketitle
%
%________________________________________________________________

\section{Introduction}

The chemical inhomogeneities (spread in the abundances of CNO, O -- Na
and Mg -- Al anticorrelation) observed in globular cluster (GC) members
from the main sequence to the red giant branch impinge on problems such as
cluster formation and early evolution (see, e.g., Kraft 1994, Gratton et al.
2001). In some clusters the peculiarities are particularly wide spread and
strong, as in M13 and NGC 6752 (Da Costa \& Cottrell 1980, Norris et al.
1981, Smith \& Norris 1993, Pilachowski et al. 1996, Kraft et al. 1997,
Sneden 2000, Gratton et al. 2001). The fact that these clusters present also
the ``anomaly" of an extended and exclusively blue horizontal branch has been
noticed by Catelan \& de Freitas Pacheco (1995), who observe that clusters
with super oxygen--poor giants have all very blue HBs.
On the other hand, it seems most likely that there is not a direct relation
between the two features (i.e., blue tail and anomalies), but rather that
they may share the same origin, at least in part.

The most common hypotheses for the origin of the chemical peculiarities are
either deep mixing in the cluster members of material nuclearly processed
in their interior, or pollution -- partial or total -- by external material
(see, e.g., Bell et al. 1981, Cottrell and Da Costa 1981, D'Antona et al.
1983, Kraft 1994, Cannon et al. 1998). The confirmation that turn--off stars
show peculiarities quite similar to those observed in giants casts severe
doubts on the deep mixing scenario (Gratton et al. 2001).

The primordial pollution hypothesis requires either contamination of
intracluster material, out of which new stars form with a chemical
composition different from a preceding star generation, or partial
contamination (superficial or deep) of pre--existing main sequence
structures. The contaminating objects are generally identified with
intermediate mass asymptotic giant branch (AGB) stars (4 -- 6 \Msun\ stars,
see D'Antona et al. 1983, Ventura et al. 2001, 2002; Thoul et al. 2002),
which evolve rapidly enough to shed their envelopes in about $10^8$ yr.

The detailed composition of these envelopes is not easy to predict, as it
depends on the uncertain physics of convection and mass loss in the
intermediate mass AGB stars, but {\it it will surely show an enrichment in
helium}, in part as a consequence of the second dredge--up, and in part due
to the third dredge--up: the helium mass fraction should be $\simgt$0.29
instead of $\sim$0.24, the cosmic value (Ventura et al. 2001)\footnote{Notice
that the value Y=0.29 for the ejecta of the most massive AGBs is
substantially conservative, as the Ventura et al. models make very
conservative assumptions on the modalities of the third dredge up.}. When
these envelopes are expelled at low velocity  by the stellar bodies during
the AGB evolution, various consequences can arise. Given the depth of the
potential well, a significant fraction of this material is not lost by a
globular cluster, unless there is a significant interaction with external
systems (e.g. another cluster, or a giant molecular cloud, or the galactic
bulge). New stars can form out of this matter, or pre--existing diffuse
material can be polluted, or low mass unevolved stars can be polluted to
various degrees.

%-------------------------------------------------------------
   \begin{figure}
   \centering
\resizebox{8.8cm}{!}{\rotatebox{0}{\includegraphics{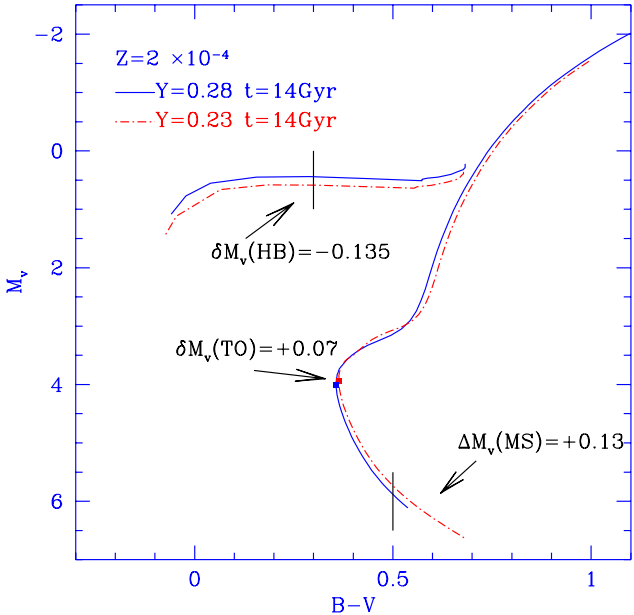}}}
      \caption{Comparison of two isochrones of age 14 Gyr, Z=$2 \times
10^{-4}$\ and different helium abundance. The ZAHBs for the two Y
values are also shown, down to a minimum HB mass of M=0.6 \Msun.}
	 \label{f1}
   \end{figure}
%-------------------------------------------------------------
   \begin{figure}
   \centering
\resizebox{8.8cm}{!}{\rotatebox{0}{\includegraphics{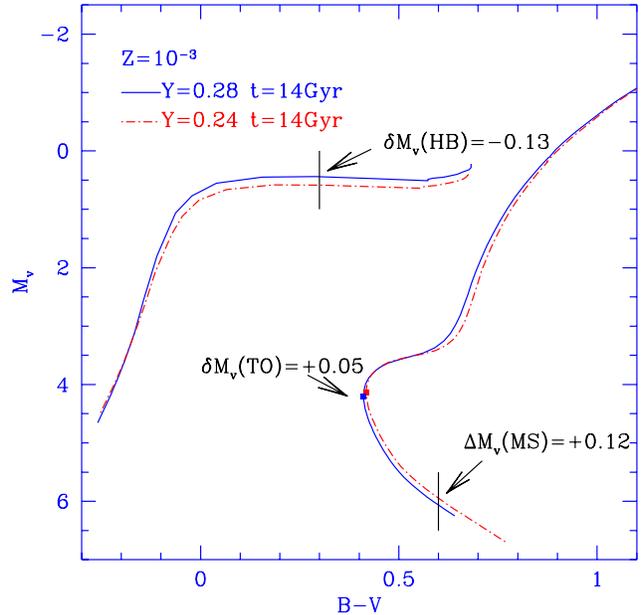}}}
      \caption{The same comparison as in Fig. \ref{f1}, for
Z=$10^{-3}$. In this case, we have extended the ZAHB down to masses as
small as M=0.5 \Msun for Y=0.24 and M=0.491 \Msun\ for Y=0.28. These
additional (bluer) models below 0.6 \Msun\ are necessary for the simulations
presented in Sect. 2.1.
	      }
	 \label{f2}
   \end{figure}

In order to evaluate the impact of a possible population having a different
chemical composition on the color magnitude diagram of a globular cluster,
in this article we compare isochrones having primordial helium abundance
\footnote{We take Y=0.24 as standard value for the
metallicity
Z=$10^{-3}$. Implicitly, we are assuming that the Big Bang value was
Y=0.23, and that there is a small increase in Y for the matter reaching a
metallicity Z=$10^{-3}$. Alternatively, we may regard Y=0.23 and Y=0.24
both as values indicative of the helium abundance emerging from the Big
Bang. The value Y=0.228$\pm$0.005 was derived by Pagel et al. (1992),
whereas today a larger value, Y=0.244$\pm$0.002 results from Izotov and
Thuan (1998) analysis of the helium emission lines in HII regions of metal
poor dwarf galaxies.}, which we set at Y=0.23 for Z=2$\times 10^{-4}$, and at
Y=0.24 for Z=$10^{-3}$, with models having a main sequence enhanced helium
Y=0.28. We compute models for all evolutionary phases and compare the
location of the main sequence (MS), turn--off (TO), red giants (RGB) and
horizontal branch (HB). The lower main sequence models, and the detailed
evolution along the RGB are considered in a forthcoming paper (Montalb\'an et
al. 2002).
In Figs. 1 and 2, the low mass cutoff of each main sequence is 0.6 \Msun.
Notice that the Y=0.28, 0.6 \Msun\ models are much more luminous (by more
than 0.4 mag) than the standard helium models. Nevertheless, the relative
locations of the standard and enhanced helium isochrones show that it is not
obvious how to discriminate between a population with uniform helium content
and one with a spread in helium, for what concerns TO and RGB evolution. On
the other hand, {\it the morphology of the HB is affected by a helium
variation, directly through the difference in the TO mass for a given age}
and we argue that a helium variation may be the main reason for the
occurrence of extreme blue tails in clusters such as NGC 6752 and M13.

This conclusion differs from the common hypothesis put forward in recent
years: a helium enrichment in the envelope as a consequence of deep mixing in
today evolving red giants has in fact been suggested as a way to obtain blue
and very blue HBs (Langer \& Hoffman 1995, Sweigart 1997). As a matter of
fact, an increased helium abundance can help in reaching a HB position bluer
than the RR Lyrae variables (at the expense of a more or less substantial
increase in luminosity), but the extremely blue locations {\it always}
require  an extreme mass loss, because the envelope mass has to be in any
case {\it smaller} than 0.01 \Msun\ (see the discussion in Caloi 2001). In
addition, Weiss et al. (2000) have shown that the helium enhancement
needed to explain the exclusively blue HB morphology would produce related
chemical inhomogeneities in red giants much stronger than observed. So an
envelope helium content larger than the cosmological one does not help in
obtaning the structures which make up the long blue tails in, e.g., M13 and
NGC 6752. We show that the situation is different if helium is larger in the
whole stellar structure, that is, if cluster stars form out of material with
enhanced helium. We considered this hypothesis at the light of the present
state of the art in stellar modelling.

\section{The models and the comparison with observations}

Appropriate evolutionary models have been computed with the code ATON2.0,
described in Ventura et al. (1998) and Mazzitelli et al. (1999). We
considered models with Y=0.28, and two values for Z: 2 $10^{-4}$ and
$10^{-3}$, to be compared with tracks with Y=0.23, Z=2 $10^{-4}$, and Y=0.24,
Z=$10^{-3}$. The increase in the helium content has been taken meaningful,
but not such as to a priori alter dramatically the currently accepted
evolution of a GC low mass star, and it roughly agrees with the expected
helium enhancement in the ejecta of massive, metal poor AGB stars. We have
built isochrones for GC ages, and transformed the luminosity and \Teff's into
observational magnitudes and colors by using the transformations by Castelli
et al. (1997).\footnote{Tables of isochrones and HB models in \Mv, B--V and
V--I are available on request. We are working to make available the models
also in other (HST) passbands, as well as a computation tool to build HB
synthetic populations, taking into account helium content spreads.}

Figures \ref{f1} and \ref{f2} show the location of the 14 Gyr isochrones in
the HR diagram, and the relevant variations in the main loci. Similar
variations hold for different ages.
The computed isochrones show that there is a small age difference for the
same TO magnitude ($\sim$0.8 Gyr for Z=10$^{-3}$, and $\sim
1$ Gyr for Z=2$\times 10^{-4}$ respectively, the larger helium the younger),
while for the same age one finds 0.05 (0.07) mag difference (the larger helium
the fainter, see Figs. \ref{f1} and \ref{f2}). The main sequence location for
the higher helium is 0.12 (0.13) mag fainter and the helium core mass at the
red giant tip is 0.008 \Msun\ smaller for the higher helium tracks (M$_{\rm
core} = 0.497$\ and 0.489 \Msun, respectively, for Z=10$^{-3}$).
The color difference between the low and high helium isochrones is
$\simlt 0.015$ mag both along the RGB and $\simeq$1 mag below the MS TO.
In a cluster with a dispersion in helium abundance, such a small difference
is certainly beyond present detection capabilities. A small population of
binaries would make even more difficult to distinguish a possible broadening
of the mean loci.

\subsection{The horizontal branch}

\begin{figure}
   \centering
\resizebox{8.8cm}{!}{\rotatebox{0}{\includegraphics{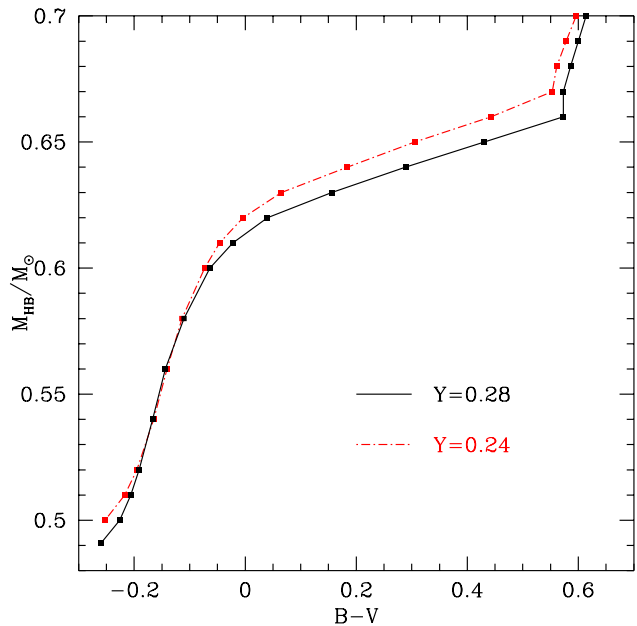}}}
      \caption{The mass versus B--V distribution in ZAHB for the two helium
contents in the Z=$10^{-3}$\ case. Formally, the Y=0.28 models are {\it
redder} that the Y=0.24 models, for the same total mass, except for the
smaller masses. However, the mass loss mechanism at the red giant branch
operates on a {\it smaller} total mass for the high helium case, so that
it is easier to achieve a smaller remnant mass and {\it bluer} stars.
}
	 \label{f4}
   \end{figure}

Table 1 shows the input parameters for the HB models, namely the helium
core mass $\rm M_{\rm c}$\ and the surface helium abundance $\rm Y_{\rm e}$.
$\rm M_{\rm c}$\ is the helium core mass at the RG tip, appropriate for ages
between 10 and 16 Gyr, according to the present models. $\rm Y_{\rm e}$\ is
increased with respect to the initial value, due to the first dredge up. The
zero age horizontal branch (ZAHB) loci are shown in Figs. \ref{f1} and
\ref{f2}, while Fig. \ref{f4} shows the HB mass versus the ZAHB B--V for the
Z=$10^{-3}$\ case.

For what concerns the actual distribution along the HB locus, the relevant
feature is that the isochrone with initial helium content of Y=0.28 reaches
the giant tip with a mass {\it smaller} of 0.05 \Msun\ than in the case with
standard helium. This is a well known result of the evolution --- see, e.g.,
Iben and Renzini (1984) --- which in our case provides an important clue to the
HB morphology of those globular clusters in which we may suspect that there
is a spread or a dichotomy in the helium abundance. Although Fig. \ref{f4}
shows that, at a given B--V color, the Y=0.28 ZAHB mass is {\it smaller} by
$\simeq 0.01 $\Msun\ than the mass for lower helium, we expect that the helium
enriched stars will occupy bluer HB loci, mainly due to the difference in the
evolving total mass.

In order to make this point more clear, we computed a few simulations of
the HB distributions expected for different assumptions on mass loss and helium
content of the sample stars. These simulations are not intended to cover in
an exhaustive way all possible cases, but simply to exemplify the main new
point of the present work.
We make the assumption that the evolving giants include two components: the
first one has initial abundance of Z=0.001 and Y=0.24; the second one has
the same metal abundance, but a uniform distribution of the initial helium
content between Y=0.24 and Y=0.28. We also ran other simulations testing the
hypothesis of a gap in the helium content distribution between the
two components. This would be the case if a new generation of stars were
born in the cluster directly from the wind ejecta of the massive AGBs.

 \begin{table}
  \caption[]{Input data for the HB models
}
     $$
	 \begin{array}{lcll}
	    \hline
	    \noalign{\smallskip}
\mathrm{Y} & \mathrm{Z} & \mathrm{Y}_e & \mathrm{M}_c/\mathrm{M}_\odot \\
	    \noalign{\smallskip}
	   \hline
	    \noalign{\smallskip}
 0.23~~~ & 2 \times 10^{-4} & 0.235~~~  &  0.508~~~ \\
 0.28    & 2 \times 10^{-4} & 0.285  &  0.497 \\
 0.24    & 10^{-3}          & 0.248  &  0.497 \\
 0.28    & 10^{-3}          & 0.2875 &  0.489 \\
	    \noalign{\smallskip}
	    \hline
	 \end{array}
     $$
 \end{table}

We assume that the individual HB masses are the result of a mass loss
process which goes on along the RGB, with a Reimers' (1997) like mass loss
law. We assume that the average mass lost in this process is $\delta$M,
with a gaussian distribution having mass dispersion $\sigma$ (in this way,
stars with identical chemical compositions may populate different part of the
ZAHBs). Further, as Reimers' mass loss rate increases with decreasing
M$_{\rm RG}$ (see the discussion in Lee et al. 1994), we also take into
account that more mass is lost on average from the helium rich (and smaller
mass) evolving giants.
We ran various simulations with different values of $\delta$M and
$\sigma$\footnote{We do not assume any explicit dependence of mass
loss on the metallicity, as we are only giving examples of the resulting
distributions for one metallicity only.}, and computed the distribution of
stars along the HB for several ages. Here we show a few examples of
the results obtained for an age of 13 Gyr. Very similar results can be
obtained for larger (smaller) ages, by decreasing (increasing) accordingly
the average mass loss.
The total number of HB stars originated from main sequence stars with Y=0.24
is taken to be 250. To these, we add from 100 to 150 stars deriving from main
sequence structures with helium content varying from 0.24 to 0.28, as
mentioned before.

\begin{figure*}
   \centering
\resizebox{14cm}{!}{\rotatebox{0}{\includegraphics{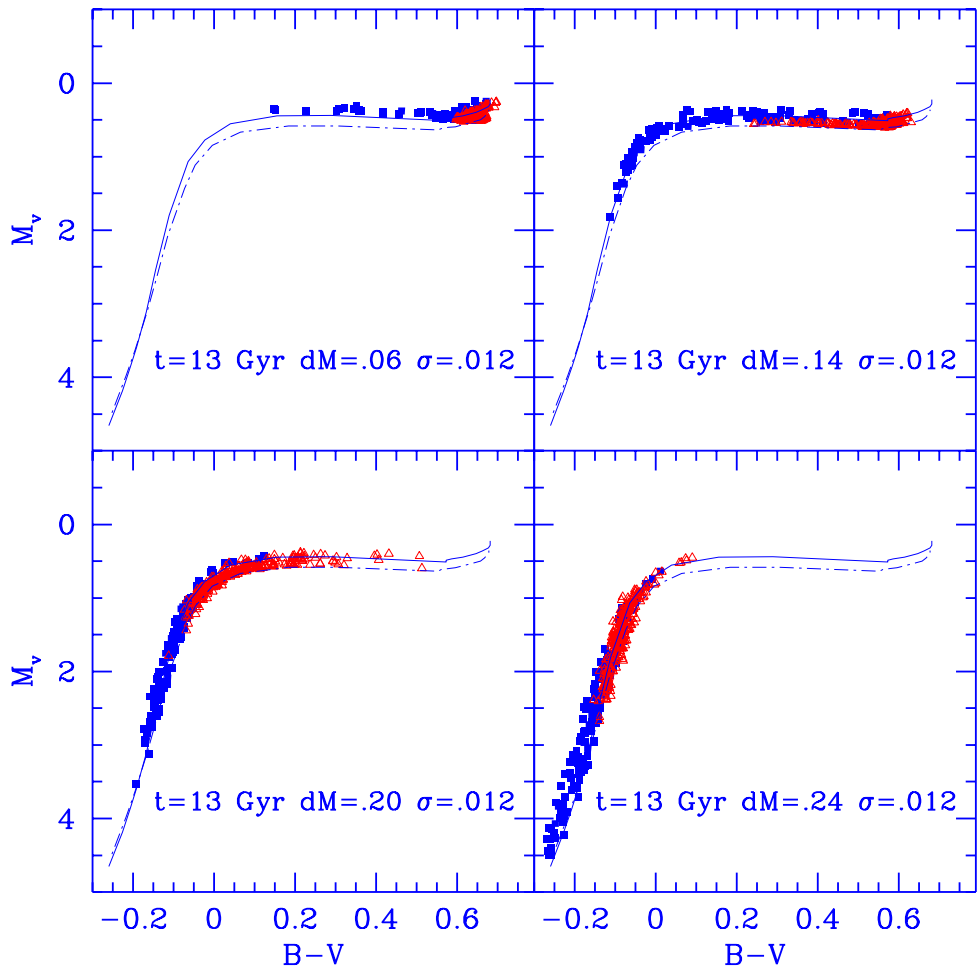}}}
      \caption{The four panels show the synthetic HB for different choices
of the average mass loss (dM in \Msun), and dispersion in mass loss
($\sigma$, in \Msun). The age assumed is 13~Gyr and the metallicity is
Z=10$^{-3}$. The simulations contain 400 stars, of which 250 have Y=0.24
(triangles) and the remaining 150 have Y randomly chosen between 0.24 and
0.28 (squares).
}
	 \label{f5}
   \end{figure*}

In the four panels of Fig. \ref{f5}, the HB members with original Y=0.24 are
indicated by open triangles, while squares indicate stars deriving from the
He--rich population described above (original Y from 0.24 to 0.28); evolution
out of the ZAHB is taken into account, together with a small spread in V
($\sigma_{\rm v}=0.005$) and in B--V ($\sigma_{\rm B-V}=0.01$), which again
simulate the impact of observational errors. Cluster members with
low helium abundance give origin to the bulk of HB distribution, while, as
expected, the high helium structures are able to populate much bluer HB
regions. This is in part a consequence of the smaller mass evolving at the TO
(up to 0.05 \Msun\ for the stars with Y=0.28), and in part because of the
extra mass loss due to the smaller mass of the evolving giants (up to 0.03
\Msun\ for Y=0.28; see Lee et al. 1994).

When the average mass loss is small, most stars would evolve on the red side
of the HB (top panels of Fig. \ref{f5}). However, stars on the red HB and
within the RR Lyrae strip computed with a distribution of initial values of
the helium abundance would be on average more luminous and yield a
substantially thicker distribution than expected in the standard (low helium)
case. For clusters more metal rich than considered here, in which the HB is
restricted to the red of the RR Lyrae gap, this feature has probably been
observed: in 47 Tuc (Briley 1997), CN--strong HB stars appear on the average
slightly more luminous (0.05 mag) and tend to be bluer, than their CN--weak
counterparts. By increasing the average mass loss in our simulations, (top
right panel in Fig. \ref{f5}) the helium rich stars begin to populate the
whole HB extension from blue to red, even assuming a relatively small
dispersion in the mass loss of $\sigma=0.012$.

   \begin{figure}
   \centering
\resizebox{8.8cm}{!}{\rotatebox{0}{\includegraphics{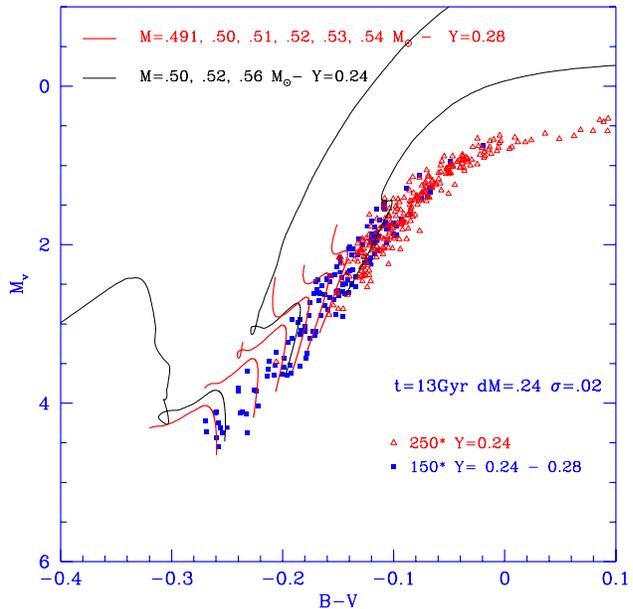}}}
      \caption{A simulation is shown together with the corresponding
evolutionary tracks.
}
	 \label{f6}
   \end{figure}

If the HB is of the type shown in the bottom left panel of Fig. \ref{f5},
the luminosity level of the HB is determined by the Y=0.24 stars, so that the
problem of predicting too high a luminosity for the HB if there is a helium
spread is not relevant: all the helium rich stars, in fact, are confined to
the blue side of the HB, due to their smaller mass and larger mass loss in
the RGB phase.

In exclusively blue HBs (bottom right of Fig. \ref{f5}), again the higher
helium structures are confined on the blue tail, where helium abundance does
not affect noticeably ZAHB position and early evolution (see, e.g., Caloi
2001).

\begin{figure}
   \centering
\resizebox{8.8cm}{!}{\rotatebox{0}{\includegraphics{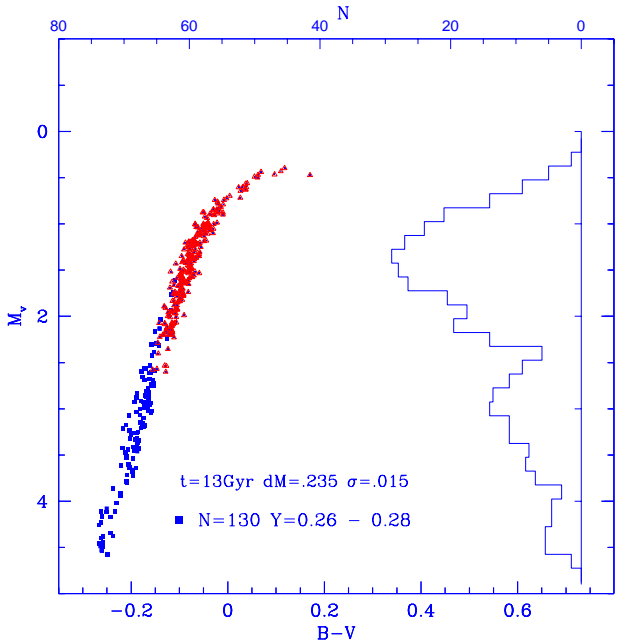}}}
      \caption{Simulation containing 380 stars, 250 of which have Y=0.24,
and 130 helium randomly chosen between 0.26 and 0.28.}
	 \label{f7}
   \end{figure}

\begin{figure}
   \centering
\resizebox{8.8cm}{!}{\rotatebox{0}{\includegraphics{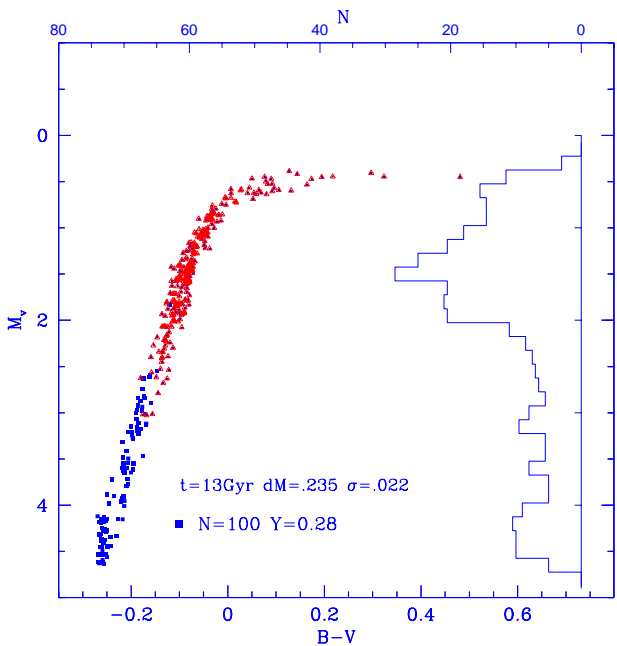}}}
      \caption{Simulation containing 350 stars, 250 of which have Y=0.24,
and 100 have Y=0.28.}
	 \label{f8}
   \end{figure}

In Fig. \ref{f6}\ we show the typical HB distribution in a simulation
including 250 stars with Y=0.24 and 150 stars with Y uniformely distributed
between 0.24 and 0.28. The corresponding tracks are also shown. The chosen
mass loss rate is such that only the helium rich stars populate the extreme
blue tail. Other experiments with different values of the mean mass loss and
mass dispersion, are shown in Figs. \ref{f7} and \ref{f8}. Here we have also
introduced a gap in the helium range of the enriched population (Y from 0.26
to 0.28 in Fig. \ref{f7} and a fixed Y=0.28 for Fig. \ref{f8}).
In Figs. \ref{f7} and \ref{f8} we also show the histogram of the
luminosity function of HB stars. We see that a dip in the distribution
appears at the magnitude corresponding to the change in the helium
abundance, and we consider this to be a meaningful feature. Other features
at larger \Mv\ do not appear to be statistically significant.

With values for the parameters largely in the accepted range, we can obtain
the morphologies of well known globular clusters with blue tails (cfr. Piotto
et al. 1999, Buonanno et al. 1986, Ferraro et al. 1998).
A very blue clump well separated from  a ``normal" blue HB (that is, derived
from giants with standard helium) can be obtained if helium in the enriched
population peaks at Y=0.28 (Fig. \ref{f8}). The bluest HB structures have a
hydrogen rich envelope of the order of 0.001 \Msun, and evolve vertically in
the CM diagram at a constant rate (as predicted by all evolutionary
computations), partially filling the gap.

\section{Conclusion}

The problem of chemical inhomogeneities in GCs is part of the wider question
of their chemical history, which appears more and more complex, and differing
from cluster to cluster. As put in evidence, e.g., by Sneden (2000), the
neutron--capture elements Ba and Eu have variable abundances without obvious
connection to overall cluster metallicity. Besides, variations in Si content
are also observed from cluster to cluster.

The fact that field stars do not show evidence of the substantial abundance
re--shuffling between CNO elements and Na, Mg and Al strongly suggests some
intra cluster processes at the origin of the phenomenon. Here we examine one
possible aspect of these (admittedly) rather vague processes.

We discussed the evolutionary properties of GC stars in which variations of
the primordial helium content are allowed. These variations are attributed to
self--pollution by an initial population of intermediate mass stars, and are
expected on the basis of the same stellar models which are thought to produce
the CNO variations among members of several GCs.

The observation of chemical peculiarities (dichotomy in the strength of CN
lines, Na and Al enhancement) in some GC giants suggested already a long time
ago (Cottrell and Da Costa 1981, Norris et al. 1981) that  intermediate mass
stars may have played a role in creating some inhomogeneity in the primordial
cloud of certain GCs. In particular, Norris et al. (1981) put in relation
chemical peculiarities with blue tails (NGC 6752), and suggested precisely
the mechanism which is at the basis of the present investigation.

We find that a spread in the helium content does not affect the morphology of
the MS, TO and RGB  in an easily observable way. The simulations in Figs.
\ref{f5} -- \ref{f8} show that also the HB luminosity level is affected in a
limited way, since only an RR Lyrae region populated exclusively by
structures with the maximum Y allowed (0.28) would stand out clearly as
peculiarly luminous. On the other hand, such an occurrence should not be
verified with standard values for the mean mass loss ($\Delta$M $\simgt 0.2$
\Msun, see Lee et al. 1994).

So age determination should not be affected in a substantial way. There may
be a small effect if we use as distance indicator the subdwarf main sequence,
since it is reasonable to assume that subdwarfs have a helium content not
affected by self--pollution.

On the other hand, the helium spread may constitute part of the ''second
parameter" problem. The presence of very blue extensions in many clusters
with predominantly blue HB, and sometimes also in HBs with RR Lyrae variables
and red members --NGC 2808 (Walker 1999), M62 (Caloi et al. 1987, Brocato
et al. 1996)-- may be related to the second parameter
problem, but this side of the question will not be considered here. The
extremely small envelope masses necessary to populate these blue tails
require mass losses, and/or ages, much larger than the average values
required to explain also exclusively, but not extended, blue HBs, in case of
uniform cosmological helium abundance. As we saw before, the smaller TO
masses (for a given age) of a helium enriched structure are crucial to
obtain the tiny envelopes of very blue HB stars.

The difficulties with the present hypothesis are mainly due to
the enrichment scenario, since it is not easy to pollute a large fraction of
original intracluster matter in a substantial way (Cannon et al. 1998).
We note that the requirement of the pollution of a large fraction of the
cluster population comes from the roughly bimodal distribution of CN
abundances and from the well developed blue tails in many clusters, if they
are related to the helium enhancement phenomenon (M13, NGC 6752).

In the self-enrichment scenarios, if the extreme blue tails can indeed be
attributed to the difference among the total masses of the stars evolving in
the RGB, due to their different helium content, the helium content must be
larger also in the center of the stars, to affect their hydrogen burning
lifetimes, and therefore we should conclude that the self--pollution has
occurred either on the gas from which the helium rich stellar generation was
formed, or it has occurred during the phases of the stars evolution in which
they were fully convective. It is also possible that the ejecta of AGBs,
collected at the center of the cluster, {\it directly form other stellar
generations.} If this is the case, some gaps in the HB stellar distributions
could reflect discontinuities in the helium content of the constituent
matter, due to the difference in the helium abundance in the ejecta of AGBs
of different mass (Ventura et al. 2002) and to some peculiar modalities of
star formation from these ejecta.

\begin{acknowledgements}
This work is supported by the Italian Ministry of University, Scientific
Research and Technology (MUIR) within the Cofin 1999 Project: ``Stellar
Dynamics and Stellar Evolution in Globular Clusters: a Challenge for the New
Astronomical Instruments" and the Cofin 2002 Project: ``New frontiers in the
study of pulsars". \end{acknowledgements}

\end{document}